# The professional trader's paradox

## Andrea Berdondini

ABSTRACT: In this article, I will present a paradox whose purpose is to draw your attention to an important topic in finance, concerning the non-independence of the financial returns (non-ergodic hypothesis). In this paradox, we have two people sitting at a table separated by a black sheet so that they cannot see each other and are playing the following game: the person we call A flip a coin and the person we'll call B tries to guess the outcome of the coin flip. At the end of the game, both people are asked to estimate the compound probability of the result obtained. The two people give two different answers, one estimates the events as independent and the other one considers the events as dependent therefore they calculate the conditional probability differently. This paradox show how the erroneous estimation of conditional probability implies a strong distortion of the forecasting skill that can lead us to bear excessive risks.

**The professional trader's paradox**

In order to explain how much danger is considering the financial returns as independent, I want to present to you this paradox. We have two people sitting at a table separated by a black sheet so that they cannot see each other and are playing the following game: the person we call A flip a coin and the person we'll call B tries to guess the outcome of the coin toss. This game lasts an arbitrary time interval and the person A has the freedom to choose how many tosses to make during the chosen time interval, the person B does not see the coin toss but can at any time, within the time interval, make a bet. When he makes a bet if he guesses the state the coin is in now, he wins. The person A decides to make a single coin flip (just at the beginning of the game) we say that the result is head, the person B decides within the same time interval to make two equal bets, betting both times on the exit of the head. The result is that B made two winning bets.

Now we ask ourselves this question: what is the correct compound probability associated with the result of this game? Let us ask this question to the person B who answers: every time I had bet I could choose between head and cross so I had a 50% chance of winning the bet; I won two bets so the compound probability is $0.5 \cdot 0.5 = 25\%$. Now let us say the same question to A the person who flip the coin, he replies: the probability is 50% I have flip the coin only one time within the defined time interval, so its prediction probability cannot be higher at 50%. The fact that the other player has made two bets has in practice only divided a bet in two is a bit 'as if to the racecourse we are made two distinct bets on the same horse on the same race, this way of acting does not increase the forecasting skill. Both answers seem more than reasonable, but as every mathematical paradox, the two answers contradict each other. At this point, will you ask yourself which of the two answers is correct?

We can resolve this paradox using the mathematical formula of the compound probability:

P(E1 ∩ E2) = P(E1 | E2) P(E2) = P(E2 | E1) P(E1).

The probability that both events (E1, E2) occur is equal to the probability that E1 occurs P(E1) multiplied by the conditional probability of E2 given E1, P(E2 | E1).

Seeing the formula, we immediately understand that the difference in response given by A and B is due to the different estimation of conditional probability P(E2 | E1). Person B estimates the conditional probability in this way P(E2 | E1) = P(E2) treating the events as completely independent,

while person A estimates the conditional probability in this other way P(E2 | E1) = 1 treating the events as completely dependent.

Which of the two answers is correct? The right answer is given by the person who has the knowledge to correctly estimate the conditional probability P(E2 | E1) and between the two players only the person that flip the coin can correctly estimate the conditional probability. Player B, on the other hand, not being able to see A that flip the coin, therefore he does not have the necessary information to estimate this probability correctly. Another way to understand this result can be found analysing the following question: what is the probability in this game of winning twice in a row by betting both times on the head?

The answer to this question is not always the same but it depends if after the first bet the person that flip the coin performs a new launch or not. If you make a new launch, the probability is $0.5 \cdot 0.5 = 25\%$ if instead as in the case of this paradox no further coin flip is performed the probability is 50%. So, in order to answer correctly, you need to know the number of launch made and this information is knows only from the person (A) that perform the coin flip and he's the only one can be correctly calculate the conditional probability.

If we bring this paradox on the financial markets, we understand that player A represent the financial instruments and player B represent the traders who try to beat the market. This gives us an extremely important result: all the traders make the same mistake, doing the same thing that player B did in this paradox. They consider their trades as completely independent of each other and this involves as we have seen, a strong distortion of the forecasting skill that can lead the traders to acquiring a false security that may lead them to bear excessive risks.

Player B, like the traders, think that the statistical information about his forecasting skill depends on his choice (I choose head instead of cross, I buy instead of selling) this is a big mistake because this statement is true only when these kinds of bets are independent of each other. In practice, this statement is true only when I place a bet by event in this case, the results are independent of each other and therefore these bets have a statistical meaning.

The problem is that in everyday life this equivalence is always respected. Therefore, our brain considers this equivalence always true so when we make trading we mistakenly consider our operations as independent despite the statistical evidence of non-independence (non-normal distribution of the results).

**Conclusion**

In this short article, I wanted to introduce one of the most important topics in finance, which concerns the non-independence of the results. Considering the financial returns as independent is equivalent to considering the financial markets stationary (ergodic hypothesis).

This hypothesis is considered by many experts not correct, on this topic have been written many articles [1], [2], [3]. What is the reason why such significant statistical evidence has been ignored, the main reason is the total lack of methods able to estimate the conditional probability P (A | B).

In my previous article [4] I have explained an innovative method used in order to evaluate a financial strategy under the condition of the market non-stationary hypothesis (non-ergodic hypothesis). This approach is based on the axiom of disorder (von Mises), this mathematical axiom applied on financial markets can be enunciated in this way:

"*Whenever we understand any kind of deterministic market process, the probability of our financial*

*operation being successful increases by more than 50%*" (Von Mises' axiom of disorder from the early 1920s).

As a consequence of this axiom given above, any correct market analysis will always tend to increase the probability of our prediction beyond the 50% mean, and this results in a consequent decrease in the probability of obtaining the same result randomly. To conclude, it follows that the parameter to be linked to the validity of a financial strategy, is not its performance but its statistical property of generating non-reproducible results in a random way.

I will demonstrate this to you with a simple example. Suppose we are playing heads or tails with a rigged coin that gives us an above-50% probability of winning (let's say it's 60%). What is the probability of losing out after 10 coin tosses? Approximately 16.6% ...and after 50 tosses? Approximately 5.7% ...and after 100 tosses? Approximately 1.7%. As you can see, the probability tends to zero, and here the rigged coin represents a financial strategy that is implementing a correct market analysis.

Now we return to the paradox that I exposed and we note how the presence of a dependence between the first and the second bet has modified the conditional probability of the second one from 0.5 to 1. This increase of the conditional probability has the consequence that the result of the second bet can be obtained randomly.

In fact, if we move the second bet randomly within the time interval from the first bet to the second one, the result is always the same because the player who flip the coin (player A) does not execute other coin tosses in this time interval. Consequently, the second bet cannot be considered to evaluate the forecast skill. Therefore, considering a system not stationary involves a reduction of the number of events to be considered for a statistical evaluation so if a data set proves to be statistically significant under the condition of stationarity of the system, the same data set may no longer be statistically significant if the system is considered non-stationary.

*E-mail address*: andrea.berdondini@libero.it